\documentclass[prl,preprint]{revtex4-1}
\usepackage{graphicx}
\usepackage{wrapfig}
\usepackage{amsmath}
\usepackage{units}
\usepackage{epstopdf}

\begin{document}
\begin{abstract}
Spontaneous transport barrier generation at the edge of a magnetically confined plasma is investigated. To this end, a model of electrostatic turbulence in three-dimensional geometry is extended to account for the impact of friction between trapped and passing particles on the radial electric field. Non-linear flux-driven simulations are carried out, and it is shown that considering the radial and temporal variations of the neoclassical friction coefficients allows for a transport barrier to be generated above a threshold of the input power.
\end{abstract}

\title{L-H transition dynamics in fluid turbulence simulations with neoclassical force balance}
\author{L. Ch\^on\'e$^{1,2}$}
\author{P. Beyer$^{1}$}
\author{Y. Sarazin$^{2}$}
\author{G. Fuhr$^{1}$}
\author{C. Bourdelle$^{2}$}
\author{S. Benkadda$^{1}$}
\affiliation{$^{1}$Aix--Marseille Universit\'e, CNRS, PIIM UMR 7345, 13397 Marseille Cedex 20, France\\$^{2}$CEA, IRFM, F-13108 Saint-Paul-lez-Durance, France.}
\maketitle

Regions of reduced energy and particle diffusion are observed in magnetic fusion devices such as tokamaks and stellarators~\cite{Stae1998,ConW2000}. These regions are named transport barriers and are equivalent to those observed in atmospheric and oceanic turbulent flows~\cite{Esle2008}. The high confinement mode or H-mode barrier, which forms at the edge of magnetic fusion devices was the earliest observed and the most studied~\cite{WagB1982}. Since then many theoretical models have been devoted to the study of this very promising confinement mode~\cite{ConW2000}. These theories share in common the fact that the potential structure that is observed in the H-mode regime, and which gives rise to a strong negative radial electric field is indeed responsible for the turbulence suppression by shear effects in the $\bf{E}\times \bf{B}$ velocity at which fluctuations are convected. On another hand, theory shows that the plasma gradients in the H-mode barrier are limited by pressure driven ballooning modes leading to relaxations of the barrier, known as Edge-Localized modes (ELMs)~\cite{WilC2006}. The transition from a regime of low confinement to one of high confinement at the edge, or L-H transition occurs when externally injecting power into the plasma and is generally followed by quasi-periodic relaxations of the barrier, which is a characteristic of the ELMs. The importance of achieving high confinement makes H-mode one of the ITER baseline scenarios, however it could be seriously  hindered by the harmful nature of ELMs to the wall components. Because of this, the understanding of the creation, control and removal of external transport barriers is of crucial importance to the success of magnetic fusion. Although the L-H transition has been widely observed and the conditions for triggering H-mode have been extensively studied experimentally, theoretical understanding of the underlying physical mechanisms remains unresolved~\cite{ConW2000,ConnorTTF}. In particular, plasma edge turbulence simulations based on first principles show self-generation of sheared flows and subsequent turbulence reduction, but no clear transition is observed~\cite{ConnorTTF}.

In this letter we present non-linear results of flux-driven resistive ballooning simulations of the plasma edge, taking into account the effect of neoclassical friction on the $\bf{E}\times \bf{B}$ flow.
It is found by means of three-dimensional (3D) simulations that competition between the neoclassical friction and zonal-flows allows for the existence of two distinct regimes depending on the imposed heat flux. These regimes correspond to a low-confinement state dominated by turbulence, and above a certain input power, a state of improved confinement with the onset of a transport barrier. Radial and temporal variations of the friction coefficients are found to have a strong impact on the dynamics of the system, so that taking them into account is necessary to obtain generation of this transport barrier. A reduced 1D model which reproduces qualitatively the 3D result is derived, and 1D simulations show intermittent bursts of turbulent flux corresponding to relaxations of the established barrier.

In the following simulations, the non-linear evolution of electrostatic resistive ballooning turbulence in 3D toroidal geometry is reproduced using the EMEDGE3D code~\cite{FuhB2008}, with the three dimensions denoted $\left(r,\theta,\phi\right)$ being the minor radius, the poloidal and toroidal angles, and their normalised counter-parts $\left(x,y,z\right)$. This code solves the following reduced MHD model, in the limit of large aspect ratios and with the slab approximation:
\begin{eqnarray}
\partial_t\nabla_\perp^2\phi+\left\{\phi,\nabla_\perp^2\phi\right\}
& = &
-\nabla_{\parallel}^2\phi-\mathrm{ G}p+\partial_xF_{neo}+\nu_{\perp}\nabla_\perp^4\phi,\nonumber\\
\label{eq:rbmw}\\
\partial_tp+\left\{\phi,p\right\}
& = &
\delta_c\mathrm{G}\phi+\chi_\parallel\nabla_\parallel^2p+\chi_\perp\nabla_\perp^2p+S.\nonumber\\
\label{eq:rbmp}
\end{eqnarray}

Equations~(\ref{eq:rbmw},\ref{eq:rbmp}) correspond respectively to the charge and energy balance, the two fields $\phi$ and $p$ being the electrostatic potential and the total pressure. $\nabla_{\parallel}$ and$\nabla_{\perp}$ are respectively the parallel and perpendicular gradients with respect the magnetic field lines and $\mathrm{G}$ is a toroidal curvature operator. $\nu_\perp$ is the classical viscosity, while $\chi_\parallel$ and $\chi_\perp$ account for parallel and perpendicular collisional heat diffusivities. $S\left(x\right)$ is a heat source term (all numerical results presented here are from flux-driven simulations). A term for poloidal flow damping which accounts for friction between trapped and circulating particles $\partial_xF_{neo}\left(\bar{\phi},\bar{p}\right)$ is added in Eq.~(\ref{eq:rbmw}) ($\bar{f}$ denotes the flux-surface average of quantity $f$). 

The system~(\ref{eq:rbmw},\ref{eq:rbmp}) is dimensionless: time is normalised to the interchange time, $\tau_{int}=\frac{\sqrt{R_0L_p}}{\sqrt{2}\tilde{c}_{S0}}$, with $\tilde{c}_{S0}$ the acoustic speed and $L_p$ the characteristic length of pressure gradient. The perpendicular length scale is the resistive ballooning length, $\xi_{bal}=\sqrt{\frac{\rho\eta_\parallel}{\tau_{int}}}\frac{L_s}{B_0}$, with the magnetic shear length $L_s$ being the parallel length scale. The fields $\phi$ and $p$ are normalised respectively to $\frac{B_0\xi_{bal}^2}{\tau_{int}}$ and $\frac{\xi_{bal}p_0}{L_p}$. Because the MHD model doesn't separate density and temperature, an assumption is made that the former is constant $n=n_0$, therefore $p=n_0T$. Furthermore, a fixed ratio between the temperatures $T_i=\epsilon_TT_e$ is also assumed. This is necessary to carry out the derivation of the neoclassical friction term.

The starting point of this reasoning is the radial force balance equation which, if we consider toroidal rotation to be negligible (generally true in the absence of torque injection), can be written thus:
\begin{eqnarray}
\partial_x\bar{\phi}+\frac{\epsilon_T}{\epsilon_T+1}\frac{\tau_{int}p_0}{\xi_{bal}L_pen_0B_0}\partial_x\bar{p}
& = &
\bar{u}_{y}.\label{eq:rfb2}
\end{eqnarray}
In the fluid model, the poloidal velocity is not normally constrained, however an expression emerges from the neoclassical theory: $\bar{u}_{y}^{neo} = \frac{\epsilon_T}{\epsilon_T+1}\frac{\tau_{int}p_0}{\xi_{bal}L_pen_0B_0}K\left(\nu_{i,*}\right)\partial_x\bar{p}$~\cite{HinH1976,HelS2005}.
The collisionality is expressed as a function of $\bar{p}$ (since $\nu_{i,*}\sim nT_i^{-2}\sim\bar{p}^{-2}$, under the assumptions mentioned above), and a heuristic closure~\cite{GiaK2002} allows for this constraint to be taken into account in the fluid model through a friction term which enforces relaxation towards this equilibrium:
\begin{eqnarray}
F_{neo}
& = &
-\mu_{neo}\left(\bar{p}\right)\left[\partial_x\bar{\phi}-K_{neo}\left(\bar{p}\right)\partial_x\bar{p}\right],
\end{eqnarray}
where $K_{neo}=\frac{\epsilon_T}{\epsilon_T+1}\frac{\tau_{int}p_0}{\xi_{bal}L_pen_0B_0}\left[K\left(\nu_{i,*}\right)-1\right]$, and $\mu_{neo}=\mu_i\left[\frac{q\left(x\right)}{\varepsilon\left(x\right)}\right]^2$, with $q\left(x\right)$ the safety factor and $\varepsilon\left(x\right)$ the inverse aspect ratio. The Hinton and Hazeltine formula is used to determine $K\left(\nu_{i,*}\right)$ for all neoclassical regimes~\cite{HinH1976}, and an approximate fit for $\mu_i$ is found in~\cite{GiaK2002}. A simple form of this friction term is obtained by considering that collisionality is constant and in the range of the plateau regime, so that $K\left(\nu_{i,*}\right)\simeq0$. One can notice that this term allows for a coupling between the flow and the pressure gradient, therefore a possible positive feedback between the two. Moreover the value toward which the rotation relaxes depends on the value of $K_{neo}$, so that radial variations of $K_{neo}$ will introduce a shear.

\begin{figure}[h]
\centering
\includegraphics[scale=0.22]{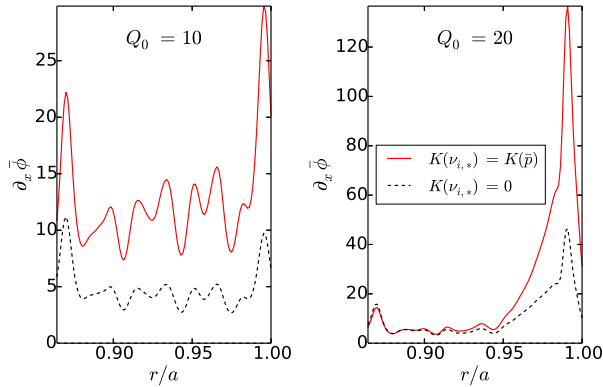}
\caption{(Color online). Comparison of the calculated mean poloidal velocity $\bar{u}_y^{FB}$ with $K\left(\nu_{i,*}\right)=0$ and $K\left(\nu_{i,*}\right)=K\left(\bar{p}\right)$, in the lower and higher range of heat flux.}
\label{fig:3D_Kneo_velo}
\end{figure}

The simulations are carried out in the range of minor radius between $0.85<r/a<1$. This main simulation domain is bounded by buffer zones where the turbulence is artificially stabilised by large $\chi_\perp$ and $\nu_\perp$.
All simulations are flux-driven by a source $S\left(x\right)$ located in the $x<x_{in}$ buffer zone, imposing the heat flux $Q_0=\int S\left(x\right)dx$. Here $x$ denotes the normalised minor radius (to $\xi_{bal}$), $x_{in}$ and $x_{out}$ are the positions of the main simulation domain's boundaries, with $x_{out}$ corresponding to $r=a$.
The safety factor is hyperbolic, between $q\left(x_{in}\right)=2.5$ and $q\left(x_{out}\right)=3.5$\,.
The set of parameters used here is representative of medium to large present tokamaks. In particular, collisionality is in the range $10^{-1}<\nu_{i,*}<10^2$ (near banana to collisional regime), which is in agreement with what is observed in L-mode at the edge~\cite{BouG2012}, and happens to be where $K\left(\nu_{i,*}\right)$ varies the most rapidly. This is with the exception of $\nu_{\perp}$ and $\chi_{\perp}$, chosen large enough to ensure damping at sub-Larmor scales. The focus of the study being the L-H transition, competition between the mean flow (here in particular due to the $F_{neo}$ term) and zonal-flows is expected. While zonal-flows generated by turbulence are included in the model, it doesn't account for certain zonal-flow saturation phenomena. In particular, it turns out that to maintain competition between both contributions to the flow, it is necessary to increase the influence of $F_{neo}$, which is done by multiplying $\mu_{neo}$ by a factor 6 (sufficient in this set of parameters).
Several simulations are done in the range of $5\le Q_0\le 30$ to study the impact of this friction on confinement.
\begin{figure}[h]
\centering
\includegraphics[scale=0.22]{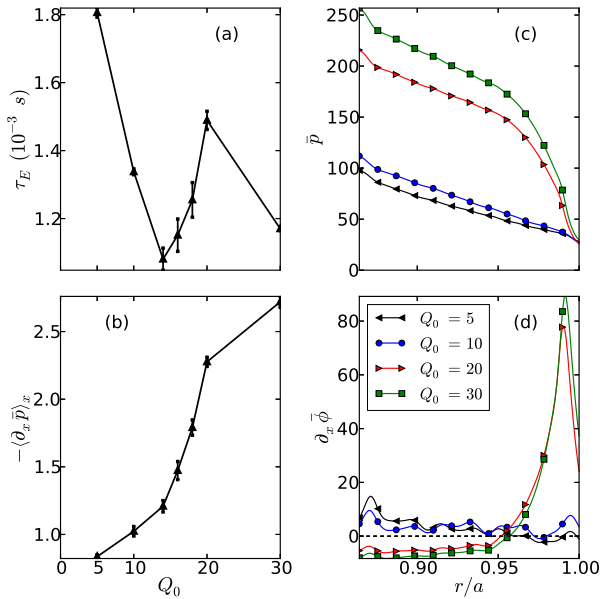}
\caption{(Color online). Evolution of the confinement efficiency as a function of the heat source amplitude in the 3D case. Left panels shows volume-averaged quantities, with error bars being the standard deviation in time. Right panels are flux-surface averaged quantities.}
\label{fig:3D_scan}
\end{figure}
The results on figure~\ref{fig:3D_scan} show the confinement deterioration expected in L-mode with increasing heat flux for $Q_0<14$, followed by a sharp increase and again a deterioration if the source is increased further. This corresponds to strong changes in the profiles of the flux-surface averaged pressure and poloidal velocity, the latter being defined as $\bar{u}_y=\partial_x\bar{\phi}$. Indeed, when the heat flux is below $Q_0=14$, the pressure profile is roughly a straight line and the poloidal velocity is low with some radial variations. Above the threshold, the poloidal velocity profile is strongly modified: in the main part of the simulation domain it stays at low amplitude and changes sign, but between $0.95<r/a<1$ it peaks strongly, generating a localised sheared flow. The stabilizing effect of sheared flows on turbulence, which is a well documented result from reduced models to gyrokinetic simulations~\cite{BigD1990,HinS1993,BeyB2005,MalD2008,StrS2013}, allows for steeper pressure gradients to be reached, giving a pedestal-like pressure profile. In the higher range of heat flux, the poloidal velocity tends towards the force-balance value, $\bar{u}_y^{FB}=K_{neo}\partial_x\bar{p}$, while it departs from it for the lower sources. Furthermore, the shape of the velocity profile shows good qualitative agreement with measurements of the radial electric field in H-mode~\cite{SauP2012,WolS2012,VieP2013}, even though the radial electric field at the LCFS is not constrained by SOL physics.
We also show that the mean value of the poloidal velocity at the peak, and consequently the associated shear, is significantly increased (here about 3 times larger) when $K_{neo}$ and its radial variations are taken into account (see~Fig.~\ref{fig:3D_Kneo_velo}). Correspondingly, the friction coefficients, as calculated from the equilibrium pressure in the code, show large changes before and after the transition.

\begin{figure}[h]
\centering
\includegraphics[scale=0.22]{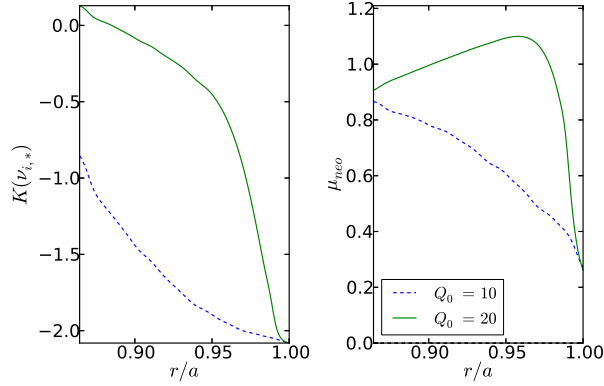}
\caption{(Color online). Profiles of $K\left(\nu_{i,*}\right)$ and $\mu_{neo}$ shown for two values of input power, before and after the transition.}
\label{fig:3D_Kneo}
\end{figure}

In particular, as shown on~Fig.~\ref{fig:3D_Kneo} (left panel)  the maximum value of $K$ goes from -1 for $Q_0<14$ (transition from collisional to plateau neoclassical regime) to 0 (plateau regime) after the transition. Moreover, after the transition the profile of $K$ is ranging from -2.1 to 0 with a strong gradient at the position of the barrier. Correspondingly, the value of $\mu_{neo}$ doubles at the position of the barrier after the transition (as illustrated in~Fig.~\ref{fig:3D_Kneo}, right panel), and shows a very sharp gradient outward from this position. This supports the fact that the radial and temporal variations of both coefficients should be taken into account, $K$ in order to allow for strong enough shear flows, and $\mu_{neo}$ in order to allow for competition between neoclassical friction and zonal-flows.


\begin{figure}[h]
\centering
\includegraphics[scale=0.22]{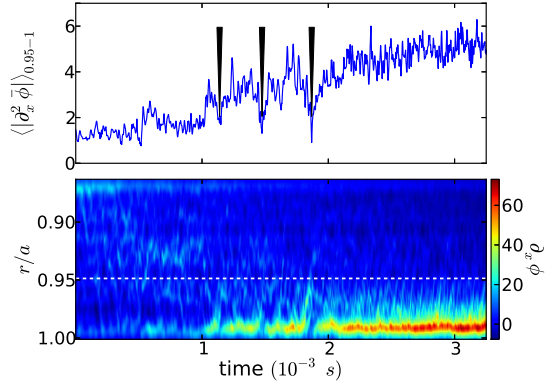}
\caption{(Color online). Time evolution of the the poloidal velocity when crossing slowly the threshold. Upper panel shows evolution of the average shearing rate in the range $0.95\le r/a\le1$, with black wedges pointing at the repeated drops during the transition. Lower panel time evolution of the poloidal velocity radial profile, with the white dashed line highlighting the position $r/a=0.95$. Here the statistically stationary phase is not shown.}
\label{fig:3D_Dithering}
\end{figure}

Smooth approach of the threshold has shown dithering of the poloidal velocity, reminiscent of the I-phase in slow L-H transitions~\cite{Zohm1994,ColS2002}.
This is clearly seen when looking at the time evolution of the poloidal velocity and the associated shearing rate, as shown on Fig.~\ref{fig:3D_Dithering}.
Before the formation of the transport barrier the velocity shear fluctuates around 1 to 2 (normalised unit). An increase to twice this value is then observed shortly after 1\,ms, soon followed by a sharp fall back to its original level. This is repeated twice, each time towards higher velocities, before a new state is reached at $t>2\,\mathrm{ms}$ and a barrier is established. As can be seen on the lower panel of Fig.~\ref{fig:3D_Dithering}, the radial maximum of the velocity corresponds to the peak observed in Fig.~\ref{fig:3D_scan}d in the case of a steady barrier.

Considering the necessity of simulations several confinement-times long to observe phenomena such as relaxations, let us now consider a 1D reduced model. Construction of this model is done using the same assumptions that led to Eqs.~(\ref{eq:rbmw},\ref{eq:rbmp}), in addition to the flute approximation $k_\parallel = 0$  to overlook the toroidal direction. If we retain only one poloidal wave-number $k$, the two fields $p$ and $\phi$ are decomposed in terms of equilibrium and fluctuating quantities thus: $f = \bar{f}+\tilde{f}\mathrm{e}^{\imath ky}+\mathrm{c.c.}$, and the following four-fields 1D system is obtained~\cite{BenB2001}:
\begin{eqnarray}
\partial_t\bar{p}
& = &
-\imath k\partial_x\left(\tilde{p}\tilde{\phi}^*-\tilde{p}^*\tilde{\phi}\right)+\chi_\perp\partial_x^2\bar{p}+S\left(x\right),\label{eq:f4fpeq}\\
\partial_t\bar{V}
& = &
\imath k\partial_x\left(\tilde{\phi}\partial_x\tilde{\phi}^*-\tilde{\phi}^*\partial_x\tilde{\phi}\right)\nonumber\\
& &
-\mu_{neo}\left(\bar{V}-K_{neo}\partial_x\bar{p}\right)+\nu_{\perp}\partial_x^2\bar{V},\label{eq:f4fveq}\\
\partial_t\tilde{p}
& = &
\imath k\left[\tilde{\phi}\left(\partial_x\bar{p}-\kappa\right)-\bar{V}\tilde{p}\right]-\alpha_p\left|\tilde{p}\right|^2\tilde{p}+\chi_\perp\partial_x^2\tilde{p},\label{eq:f4fpk}\\
\partial_t\tilde{\phi}
& = &
\imath\left(\frac{g}{k}\frac{\tilde{p}}{\bar{p}}-k\bar{V}\tilde{\phi}\right)-\alpha_\phi\left|\tilde{\phi}\right|^2\tilde{\phi}+\nu_\perp\partial_x^2\tilde{\phi},\label{eq:f4ffk}
\end{eqnarray}

with the equilibrium poloidal velocity $\bar{V}=\partial_x\bar{\phi}$. The $\alpha_f\left|\tilde{f}\right|^2\tilde{f}$ terms account for saturation via mode coupling. Here $t$ is normalised to $\frac{1}{\omega_S}=\frac{m_i}{eB_0}$, $x$ to $\rho_S=\frac{\sqrt{m_ik_BT_e}}{eB_0}$.

\begin{figure}[h]
\centering
\includegraphics[scale=0.22]{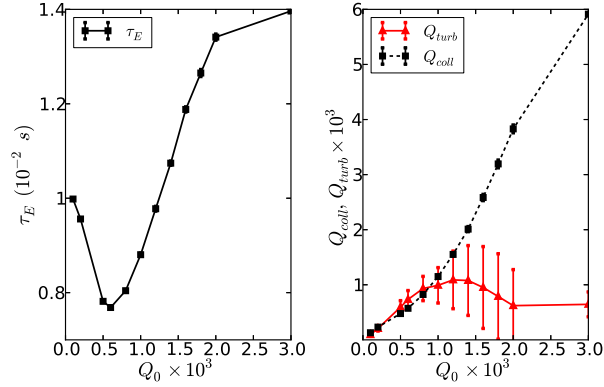}
\caption{(Color online). Evolution of the confinement efficiency as a function of the heat source amplitude in the 1D case.}
\label{fig:1D_Scan}
\end{figure}

\begin{figure}[h]
\centering
\includegraphics[scale=0.22]{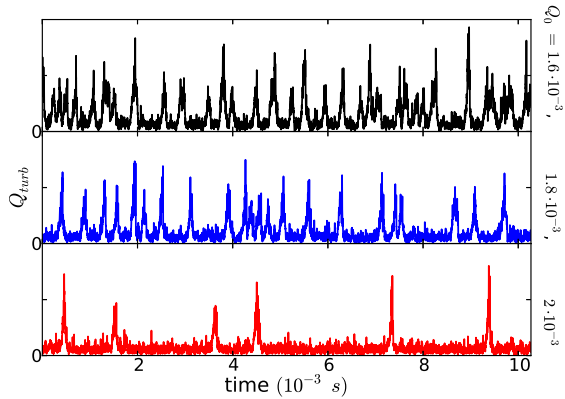}
\caption{(Color online). Time evolution of the turbulent flux in the presence of a barrier for different input powers, in the 1D case. The panels are in order of increasing power. }
\label{fig:1D_ELM}
\end{figure}

In this case, partial stabilisation of the turbulence is achieved above a certain threshold of the injected power, as illustrated on Fig.~\ref{fig:1D_Scan}, showing that this reduced model still contains the minimal elements to reproduce this behaviour. In the parameter range considered so far, it turns out that for low fluxes, the collisional and turbulent fluxes are of the same order of magnitude (Fig.~\ref{fig:1D_Scan}, right panel). These results show particularly interesting dynamics of the system once the turbulence level is strongly reduced: here turbulence is not steadily suppressed but shows instead quasi-periodic bursts. Interestingly the pseudo-period increases with the injected power (see~Fig.~\ref{fig:1D_ELM}). This behaviour bears similarities with type-III ELMs, which were already suggested to be governed by the resistive ballooning instability~\cite{FuhB2008,BeyB2005,BeyB2007}.

In conclusion, 1D and 3D fluid, non-linear flux-driven simulations of edge turbulence including self-consistent friction between passing and trapped particles have shown the existence of two distinct regimes depending on the imposed heat flux. At low heat flux, the poloidal velocity is dominated by zonal-flows, and only poor confinement is achieved. When the input power exceeds a certain threshold, the effect of friction takes over and the poloidal velocity rises sharply near the LCFS. This is associated with strong velocity shear, which governs the reduction of the turbulent transport, and the subsequent better confinement. In addition, oscillations of the poloidal velocity are observed in 3D simulations when approaching the threshold slowly, which is reminiscent of the limit-cycle oscillations in L-I-H transition experiments~\cite{Zohm1994,ColS2002,SchZ2012}. Long 1D simulations have shown quasi-periodic relaxations of the transport barrier with a period increasing with the input power, as observed for type-III ELMs.

The authors acknowledge fruitful discussions with X.~Garbet and Y.~Camenen.
This work is supported by the French National Research Agency, project ANR-2010-BLAN-940-01. Computations have been performed at the M\'esocentre d'Aix-Marseille Universit\'e.

\end{document}